\documentclass[conference]{IEEEtran}
\IEEEoverridecommandlockouts
\usepackage{cite}
\usepackage{amsmath,amssymb,amsfonts}
\usepackage{algorithmic}
\usepackage{graphicx}
\usepackage{textcomp}
\usepackage{xcolor}
\usepackage{censor}
\usepackage{multirow}
\usepackage[hidelinks]{hyperref}
\def\BibTeX{{\rm B\kern-.05em{\sc i\kern-.025em b}\kern-.08em
    T\kern-.1667em\lower.7ex\hbox{E}\kern-.125emX}}
\begin{document}

\title{WIP: Exploring the Value of a Debugging Cheat Sheet and Mini Lecture in Improving Undergraduate Debugging Skills and Mindset\\
\thanks{This work was supported by the National Science Foundation Award EES-2321255.}
}

\author{\IEEEauthorblockN{Andrew Ash and John Hu}
\IEEEauthorblockA{\textit{School of Electrical and Computer Engineering, Oklahoma State University} \\
Stillwater, OK, 78074, USA\\
Emails: \{andrew.ash, john.hu\}@okstate.edu}
}

\maketitle

\renewcommand{\thefootnote}{}
\footnotetext{%
© 2025 IEEE. Personal use of this material is permitted. Permission from IEEE must be obtained for all other uses, in any current or future media, including reprinting/republishing this material for advertising or promotional purposes, creating new collective works, for resale or redistribution to servers or lists, or reuse of any copyrighted component of this work in other works.

This is the accepted version of a paper accepted for presentation at the 2025 IEEE Frontiers in Education Conference (FIE). The final version will be available via IEEE Xplore at: \url{https://ieeexplore.ieee.org}
}

\begin{abstract}
This work-in-progress research paper explores the efficacy of a small-scale microelectronics debugging education intervention utilizing quasi-experimental design in an introductory microelectronics course for third-year electrical and computer engineering (ECE) students. In the first semester of research, the experimental group attended a debugging ``mini lecture" covering two common sources of circuit error and received a debugging cheat sheet with recommendations for testing and hypothesis formation. Across three debugging problems, students in the experimental group were faster by an average of 1:43 and had a 7\% higher success rate than the control group. Both groups demonstrated a strong general growth mindset while the experimental group also displayed a shift in their debugging mindset by perceiving a greater value towards debugging. Though these differences are not yet statistically significant, the pilot results indicate that a mini-lecture and debugging cheat sheet are steps in the right direction toward improving students' readiness for debugging in the workplace.
\end{abstract}


\section{Introduction}
The unpredictable nature of debugging in the semiconductor industry leads some engineers to dedicate over 40\% of their time to debugging \cite{mutschler_debug_2018}. The high cost of time spent in debugging has earned it the nickname ``the schedule killer" \cite{bailey_debug_2021}. Despite its critical importance in industry, debugging is infrequently taught in undergraduate electrical and computer engineering (ECE) curricula \cite{duwe_defining_2022, radu_integrating_2008}. Students are often expected to learn this vital skill through unguided debugging in labs and projects.

Past research has led to the development of hardware debugging curriculum at the undergraduate level \cite{duwe_defining_2022, radu_integrating_2008, nagvajara_design-for-debug_2007, sitik_microcontroller-based_2013}. However, as noted by Duwe \textit{et al.} \cite{duwe_defining_2022}, there is not a curriculum with qualitative and quantitative instrumentation on this topic. To fill this need in undergraduate ECE, we are developing a new circuit debugging curriculum. We apply a quasi-experimental design and collect primarily quantitative and some qualitative data on students’ debugging skills and mindset from experimental and control groups. This work presents the results of a limited debugging skills intervention and explores the following research questions:
\begin{enumerate}
    \item[] \textbf{RQ1:} To what extent does our intervention improve students’ debugging skills?
    \item[] \textbf{RQ2:} To what extent does our intervention affect students’ debugging mindset?
\end{enumerate}

The rest of the paper is organized as follows. Section II defines debugging skills and mindset, and provides background in assessment in debugging curriculum.  Section III examines the targeted debugging skills intervention. Sections IV and V report the difference in debugging skills and mindsets between control and experimental groups. Section VI shares the students' qualitative feedback. Section VII outlines future work in the project. Section VIII concludes the paper.

\section{Debugging Education Background}
\subsection{Defining Debugging Skills and Mindset}
The debugging mindset proposed by Duwe \textit{et al.} \cite{duwe_defining_2022} consists of ``(1) observable behaviors that reflect effective debugging practice and (2) internal beliefs or worldviews that guide these behaviors" with seven elements between the two parts. While circuit debugging is a specialized skill set, it falls within troubleshooting \cite{katz_debugging_1987}. Jonassen and Hung analyze the tasks a novice troubleshooter must learn to achieve an expert level of troubleshooting capability in any domain \cite{jonassen_learning_2006}: ``construct problem space," ``identify fault symptoms," ``diagnose faults," ``generate and verify solutions," and ``remember experience." Applying this ``Cognitive Model of Troubleshooting" to circuit debugging, we view students' debugging skills as their ability to \textbf{(1) identify the root cause(s)} of unexpected circuit behaviors and \textbf{(2) take corrective actions} to restore the circuit to the desired state. Three elements of Duwe's debugging mindset fall within this view of debugging skills (a holistic/system-level view, using systemic processes, and a commitment to rigor). In this work we seek to measure the other four elements of Duwe's definition of a debugging mindset:
\begin{itemize}
    \item A domain-specific growth mindset
    \item A testing mindset
    \item Productive emotions (i.e. curiosity instead of frustration)
    \item Associating value with debugging
\end{itemize}

\subsection{Assessments in Debugging Education}
Few works in literature examine circuit debugging education. One undergraduate digital hardware design course included a design for debugging focus \cite{nagvajara_design-for-debug_2007} emphasizing verification and debugging tasks, but did not include any form of assessment specific to debugging \cite{radu_integrating_2008}. In a series of studies implementing hardware/software debugging curriculum in high school classrooms, Fields \textit{et al.} used classroom observations, self-reflections, and interviews to qualitatively analyze improvements in students' debugging capabilities \cite{fields_debugging_2021}. More recently, their research has included pre- and post-surveys where students were given images and a description of buggy e-textile projects; the quantity and specificity of root causes students brainstormed revealed qualitative improvements in their debugging capabilities \cite{morales-navarro_failure_2023}. The most relevant debugging education assessments come from Duwe \textit{et al.} \cite{duwe_defining_2022}. Students in two undergraduate computer engineering courses practiced various aspects of debugging embedded systems and computer architectures in lectures, lab exercises, self-reflections, and other assignments. Student data and instructor observations were analyzed for evidence of developing a debugging mindset. The research team identified the debugging mindset was emerging among students; although, the authors note that ``analysis was not designed to identify causal links between course pedagogy and the development of debugging mindsets among students" and recommended creating ``qualitative and quantitative instrumentation to support educators and researchers in identifying the debugging mindset among engineering students" as part of future research \cite{duwe_defining_2022}.

\section{Intervention Design}
Our intervention is a novel combination of holistic and domain-specific debugging education focused on microelectronics (rather than embedded systems or hardware/software interfaces), along with assessments that combine qualitative and quantitative data on students' debugging skills.

\textbf{\textit{Participants}} were undergraduate students in a third-year microelectronics course focused on semiconductor devices and amplifiers \cite{sedra_microelectronic_2015}. Students were informed about the voluntary nature of data collection and could request that their data be excluded from research. \textbf{\textit{Utilizing a quasi-experimental design}}, the control group consisted of two lab sections with 37 participants, while the experimental group included one lab section with 13 participants. Before the semester began, a lab section was selected as the experimental group. Students selected a lab based on their schedule; there were no special criteria to place a student in the experimental group. All participants attended the same lectures (except for the intervention-specific lecture), received identical assignments, used the same lab environment, and were graded using the same metrics.

\begin{figure}[tb]
\centerline{\includegraphics[width=\columnwidth]{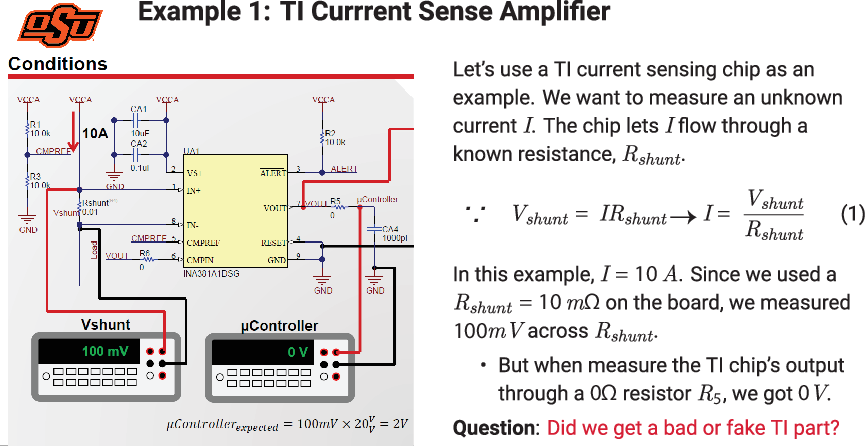}}
\caption{One slide from the mini lecture depicting connectivity issues caused by cold solder joints \cite{noauthor_precision_2021}.}
\label{mini_lecture}
\end{figure}

\begin{figure}[tb]
\centerline{\includegraphics[width=\columnwidth]{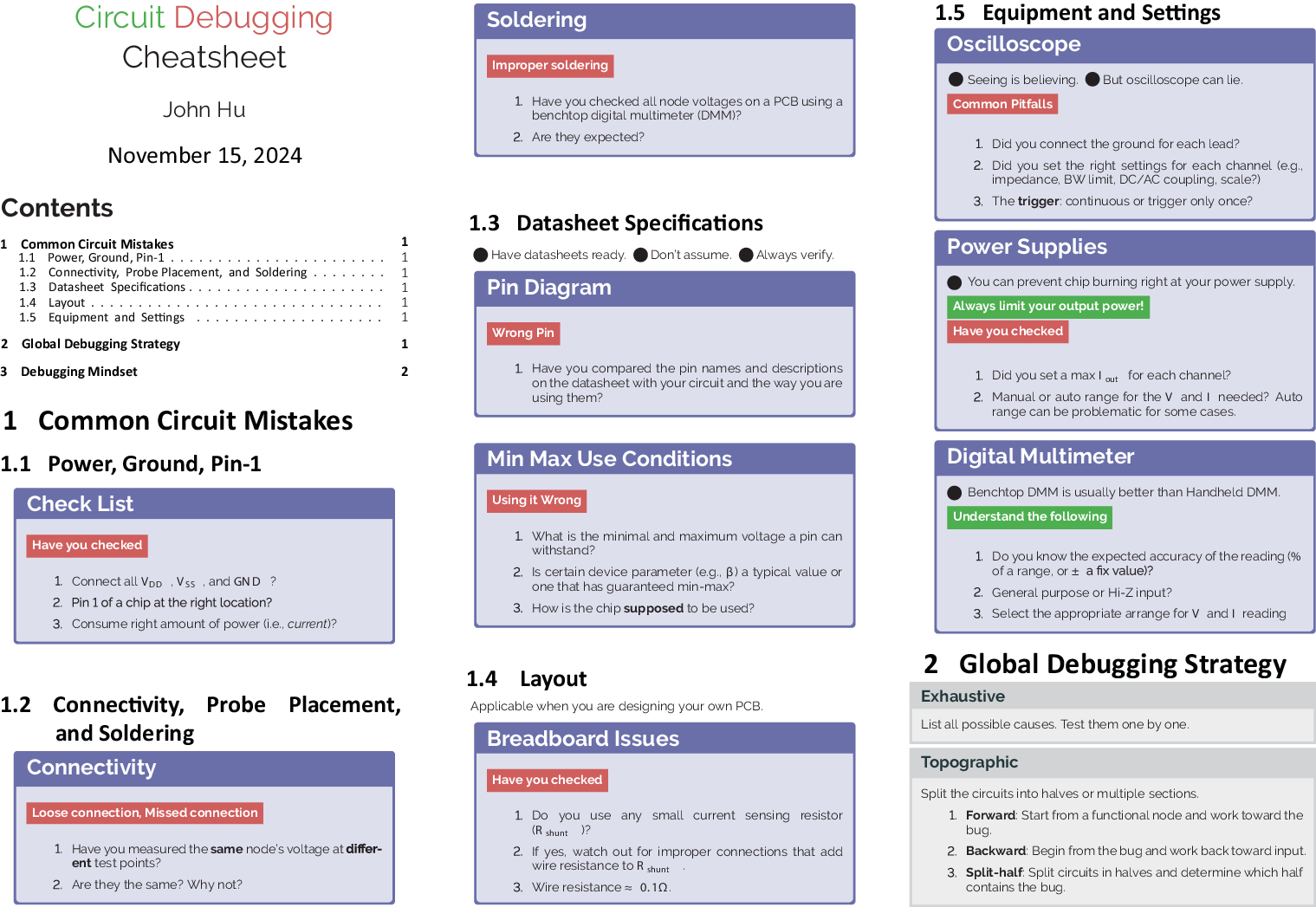}}
\caption{A page from our debugging cheat sheet depicting common bugs and bug location strategies. The cheat sheet is available on GitHub.}
\label{cheat_sheet}
\end{figure}

The first stage of our \textbf{\textit{iterative intervention design}} consisted of a mini lecture where students were introduced to two common hardware bugs and a debugging cheat sheet. The slides were adapted from circuit debugging resources developed by Texas Instruments that focused on connectivity issues as shown in Fig.~\ref{mini_lecture} and overlooking device specifications \cite{noauthor_precision_2021}. Our debugging cheat sheet, see Fig.~\ref{cheat_sheet}, lists common hardware bugs with questions to guide a bug location strategy and identify root causes and is available on GitHub \cite{debugging_cheatsheet}. Students were informed that there were no restrictions on resources used in the debugging exam, and were encouraged to use the cheat sheet to aid their debugging process.

\textbf{\textit{Quantitative data}} comes from our debugging exam's success rates, completion times, and our Likert-scaled debugging mindset survey. \textbf{\textit{Qualitative data}} includes student feedback on the exam, free response questions on the survey, and researcher observations during the exam. We analyzed qualitative data using an ongoing constant comparative method \cite{charmaz_constructing_2006}. 

\begin{table*}[tbh]
\caption{Student Success Rates and Completion Times}
\begin{center}
\begin{tabular}{|c||c|c||c|c||c|}
\hline
&\multicolumn{2}{|c||}{\textbf{Control}}&\multicolumn{2}{|c||}{\textbf{Experimental}}&\\
\hline
\textbf{Problem}&\textbf{Success Rate}&\textbf{Avg. Success Time}&\textbf{Success Rate}&\textbf{Avg. Success Time}&\textbf{Success Rate P-Value}\\
\hline
OR&7/13 (54\%)&27:00&3/5 (60\%)&25:20&0.8139\\
\hline
10X&4/7 (57\%)&11:31&2/2 (100\%)&16:30&0.2568\\
\hline
FB&7/10 (70\%)&15:53&3/5 (60\%)&10:20&0.6985\\
\hline
PMOS&7/7 (100\%)&13:58&1/1 (100\%)&16:00&N/A\textsuperscript{*}\\
\hline
Total&25/37 (68\%)&17:46&9/13 (69\%)&17:20&0.9119\\
\hline
Total (without PMOS)&18/30 (60\%)&19:14&8/12 (67\%)&17:31&0.6877\\
\hline
\multicolumn{6}{l}{\textsuperscript{*} Pooled proportion of 1 leads to division by 0}\\
\end{tabular}
\label{success_rates}
\end{center}
\end{table*}

\section{Debugging Skill Changes}
\begin{figure*}[tbh]
\centerline{\includegraphics[width=\textwidth]{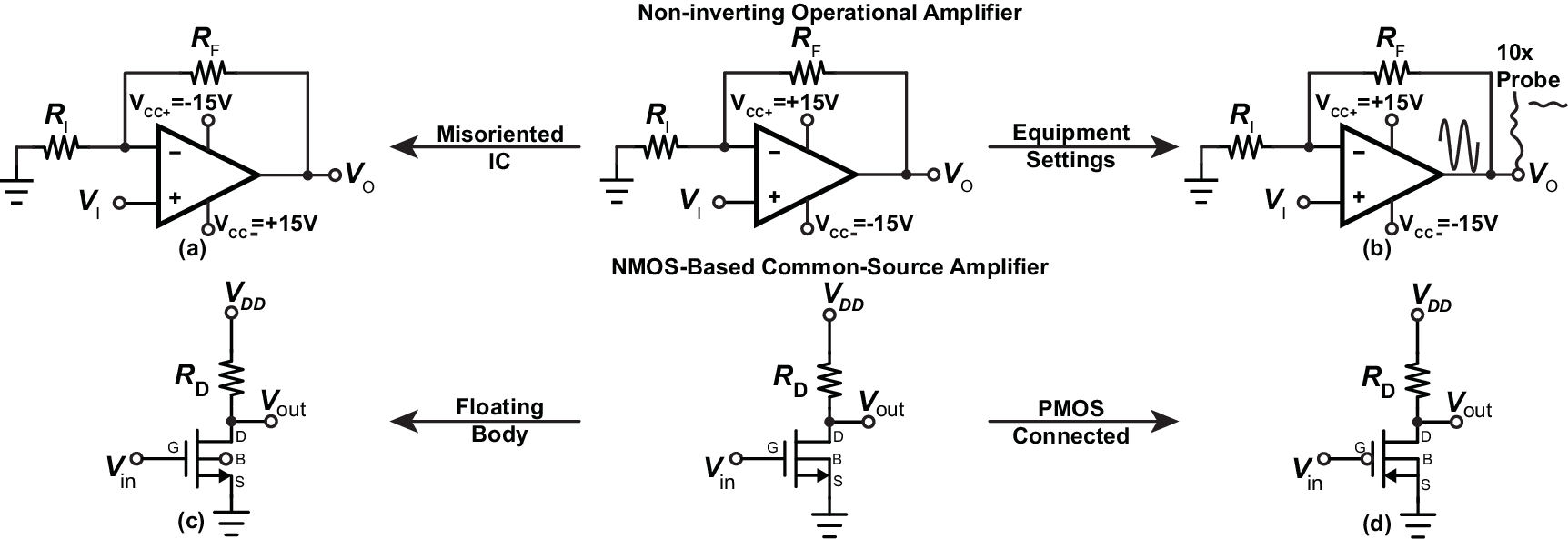}}
\caption{The four buggy circuit problems in the debugging exam. Bugs are based on a non-inverting op-amp and a common-source amplifier.}
\label{exam_circuits}
\end{figure*}

Students completed one of four problems shown in Fig.~\ref{exam_circuits}: (a) a non-inverting operational amplifier (op-amp) misoriented on a PCB to swap the +15V and -15V connections, leading to an output stuck near 0V; (b) a functional non-inverting op-amp on a PCB with an oscilloscope probe measuring the output set to 10x mode while expecting a 1x mode input, attenuating the measured gain; (c) a common-source NMOS-based amplifier with a floating body connection on a bread board, decreasing the expected gain; and (d) a common-source NMOS-based amplifier implemented with a PMOS on a bread board, also reducing the gain. Students use these circuits in the course labs, ensuring they are familiar with with the equipment and circuit theory required for the debugging analysis.

Table~\ref{success_rates} summarizes the results of the debugging exam. The average success time only considers the submission times of students who successfully identified and fixed the bug. The misoriented IC problem proved the most challenging for both groups of students, with success rates of 54\% (7/13) and 60\% (3/5) and submission times of 27:00 and 25:20 for the control and experimental groups, respectively.

At 57\% (4/7), the control group had approximately the same success rate for the oscilloscope settings problem, but the successful students submitted their results faster with an average time of 11:31. The difference may be in part because the fix for this problem only requires flipping a switch on a probe. The students assigned to the 10x probe problem in the experimental group completed the exam with a 100\%  success rate (2/2) and an average time of 16:30.

The floating body bug had higher success rates at 70\% (7/10) in 15:53 and 60\% (3/5) in 10:20 for the control and experimental groups. These faster success times may be caused by a combination of reduced visual complexity compared to the misoriented op-amp and the simple single-wire fix connecting body and ground to achieve the expected gain.

Student feedback and researcher observations indicated that the PMOS problem was too easy due to clear differences between the expected and buggy circuit implementations and the recency of the common-source amplifier lab. The 100\% success rate of this problem reduces the discriminative power of the exam, artificially inflating the success rate of both groups. For this reason, the research team excluded the PMOS problem data before analyzing the average performance across all problems. The control group's composite success rate drops to 60\% with an average time of 19:14, while the experimental group's success rate becomes 67\% (8/12) at 17:31. 

The individual and composite P-values indicate that differences between the groups are not yet statistically significant. As data collection continues in upcoming semesters, a larger sample size should reveal the full extent of improvements due to the new curriculum. However, there are emerging trends that indicate our curriculum can improve students' debugging skills. For the misoriented IC and floating body circuits, students in the experimental group successfully debugged the circuit faster on average. In addition, the success rates for the misoriented IC and 10x probe bugs are higher.

\section{Debugging Mindset Shift}
Two weeks after the debugging exam, students completed a survey to identify differences in mindset between the control and experimental groups. The survey had three parts: validated growth mindset questions \cite{stanford_university_growth_nodate}, new debugging mindset questions, and open-ended feedback about the debugging exam and curriculum. The two mindset sections used a 6-point Likert scale (strongly agree (1) to strongly disagree (6)), while the final section used text boxes for free expression. One student from the control group did not complete the survey, leading to 36 control and 13 experimental responses.

\subsection{Growth Mindset}
Table~\ref{growth_mindset} summarizes the results of the growth mindset questions. S1, S2, and S3 are taken from \cite{stanford_university_growth_nodate} as follows: ``You have a certain amount of intelligence, and you can’t really do much to change it"; ``Your intelligence is something about you that you can’t change very much"; ``You can learn new things, but you can’t really change your basic intelligence". Across the three statements, students in the control group were between a 4 (mostly disagree) and 5 (disgree) with standard deviations between 1.34 and 1.55, while the experimental group averaged at least 5 (disagree) across the three statements with standard deviations between 0.73 and 0.83. The p-values show that there is not yet a statistically significant difference between the populations, likely due to the limited sample size.

\begin{table}[tbh]
\caption{Growth Mindset Post Debugging Exam}
\begin{center}
\begin{tabular}{|c||c|c||c|c||c|}
\hline
&\multicolumn{2}{|c||}{\textbf{Control (36)}}&\multicolumn{2}{|c||}{\textbf{Experimental (13)}}&\\
\hline
\textbf{Statement}&\textbf{Avg.}&\textbf{St. Dev.}&\textbf{Avg.}&\textbf{St. Dev.}&\textbf{P-Value}\\
\hline
S1&4.47&1.34&5.08&0.73&0.2481\\
\hline
S2&4.58&1.38&5.08&0.83&0.4084\\
\hline
S3&4.22&1.55&5.00&0.78&0.2088\\
\hline
\end{tabular}
\label{growth_mindset}
\end{center}
\end{table}

These results only reflect the two populations' mindsets after the exam. Although we do not have pre-debugging lecture survey results to conclusively demonstrate that differences in growth mindset are caused at least in part by the new curriculum, the results do indicate a small difference in growth mindset between the two populations. The two groups had identical course experiences aside from the mini lecture and debugging cheat sheet, which may indicate that even the small-scale intervention in our first semester had a positive impact on our students. The survey results show that the students in our university's ECE program adopt a growth mindset, as both populations tended to ``mostly disagree" or ``disagree" with statements that intelligence is fixed. This trend holds promise that students will be receptive to an expanded debugging curriculum that targets developing a debugging mindset.

\subsection{Debugging Mindset}
Table~\ref{debugging_mindset} summarizes the results of the debugging mindset questions. The statements map to four elements of the debugging mindset from Duwe \textit{et al.} \cite{duwe_defining_2022}. Across all four statements, there is little difference between the control and experimental groups. Given the minimal debugging training students experienced, this is reasonable. That being said, for S2, S3, and S4 the experimental group shows a trend \textit{may} be developing that students with debugging training are more likely to scrutinize untested components, associate positive emotions with debugging, and see the value of debugging. Although there is no difference between the two group in their view of S1, it is encouraging to see that even without a targeted debugging curriculum students tend to adopt a growth mindset toward debugging skills; this also indicates that students will likely be receptive to expanded debugging curriculum in future semesters. The p-values indicate that a larger sample size is needed to draw more certain conclusions.

\begin{table*}[tbh]
\caption{Debugging Mindset Post Debugging Exam}
\begin{center}
\begin{tabular}{|l||c|c||c|c||c|}
\hline
&\multicolumn{2}{|c||}{\textbf{Control (36)}}&\multicolumn{2}{|c||}{\textbf{Experimental (13)}}&\\
\hline
\multicolumn{1}{|c||}{\textbf{Statement}}&\textbf{Avg.}&\textbf{St. Dev.}&\textbf{Avg.}&\textbf{St. Dev.}&\textbf{P-Value}\\
\hline
S1: ``Debugging is a skill that can be learned."&1.53&0.73&1.54&0.63&0.8474\\
\hline
S2: ``If a component wasn't tested, it doesn't work (there can be bugs)."&3.22&1.29&3.00&1.24&0.6753\\
\hline
S3: ``I feel comfortable working with bugs instead of frustrated or afraid."&2.64&1.18&2.54&1.01&0.8209\\
\hline
S4: ``Debugging is crucial and valuable for new product development in semiconductors."&1.61&0.86&1.23&0.58&0.2088\\
\hline
\end{tabular}
\label{debugging_mindset}
\end{center}
\end{table*}

\section{Qualitative Feedback}
The final section of the survey allowed students to give feedback on the efficacy of the cheat sheet on their debugging process and to offer suggestions for what topics to include in an expanded debugging curriculum within the course. Twenty students provided responses to these open-ended questions. Students' perceptions of the debugging cheat sheet were mixed; four students found the reminders helpful, two others felt they did not have enough time before the exam to really absorb the content of the sheet, and all other respondents either did not have or use the sheet. This feedback indicates students may find the debugging cheat sheet more valuable if they are introduced to it earlier in the semester to gain practice with testing for and identifying common errors.

Answers to the final question offered curriculum design insights. Table~\ref{student_responses} shows exemplar quotes of themes identified during analysis. First, students want to learn a guided way of developing hypotheses to investigate during debugging. This stage in troubleshooting proves particularly difficult \cite{alaboudi_using_2020, alaboudi_hypothesizer_2023, ma_how_2024}, guiding students in this critical debugging stage could greatly improve their skills. Students also requested guided debugging tasks throughout the semester to begin developing an intuition for debugging and gain a better understanding of what is expected in circuit debugging; one student even noting that a whole course dedicated to the basics of debugging would be desirable. Finally, students were most interested in learning a systematic process for debugging. Students' feedback consistently included thoughts about the unguided nature of their current processes, and reflecting on how debugging might be a more approachable task if they were familiar with ``the basics" or a general way of approaching debugging that would be relevant outside of the current class. It is encouraging to see the students motivated by the notion of applying their debugging skills in a generalizable sense, necessary for the unknown debugging tasks they may encounter in the workplace.

\begin{table*}[t]
\caption{Student Suggestions for a Circuit Debugging Curriculum}
\begin{center}
\begin{tabular}{ |c|c| }
\hline
\textbf{Requested Element}&\textbf{Verbatim Quotes from Students}\\
\hline
&``Honestly there is not a very specific way to teach how to debug since troubleshooting is a skill that requires a lot\\
Hypothesis generation training&of experience and intuition. I would say that asking the students the question `what can be wrong with this circuit?'\\
&might help students to get started with the basics to troubleshoot and how to debug."\\
\hline
\multirow{4}{*}{Guided debugging tasks}&``I feel like what I would need to develop is stronger intuition around electronics, which could be taught\\
&by having students guided through debugging tasks by a TA or professor on a common circuit or one-on-one."\\
&``We should have one on one time or actually work debugging in the lab if we are to learn debugging."\\
&``I wish there was a full chapter on debugging as well as a lab for it, if not an entire class dedicated to debugging."\\
\hline
&``I wish we would have gotten more information and tips on how to debug in general."\\
Systematic debugging process&``I wish that we had more information on what components to look at when debugging...\\
&showing what an incorrect output could look like and where to start backtracking from to find possible issues."\\
\hline
\end{tabular}
\label{student_responses}
\end{center}
\end{table*}

\section{Future Work}
In the next iteration, we will include more debugging lectures. We will improve our survey with three statements for each element of the debugging mindset to ensure we measure the intended constructs \cite{devellis_scale_2017} and we will validate the survey in accordance with American Educational Research Association standards \cite{american_educational_research_association_standards_2014}. Combining these curriculum and assessment improvements with student suggestions will improve the effectiveness of our research. Continued data collection will also increase our sample size to approach statistical significance.

\section{Conclusion}
This paper presents the preliminary results of introducing circuit debugging curriculum into an undergraduate electrical and computer engineering course. The debugging education intervention, if effective, may also be useful to other engineering disciplines to improve their students’ troubleshooting capabilities for complex engineering systems. While there is still work to be done, the preliminary results are promising. With one mini lecture and access to a debugging cheat sheet during the exam, students in the experimental group had 7\% higher average success rate across three problems and submitted their successful solutions 1:43 faster. In addition, they more strongly embraced three aspects of the debugging mindset. Further data collection is required as these differences are not yet statistically significant. However, students in both groups tended to adopt a general growth mindset and a domain-specific debugging mindset; as we expand the debugging curriculum in the next stage of our research, this tendency should help students improve their debugging skills and embrace a more complete debugging mindset.


\end{document}